\documentclass{iaus}
\usepackage{graphics}

\title[Lithium in Very Metal-poor Stars]
{New Keck Observations of Lithium in Very Metal-poor Stars}

\author[Boesgaard, Novicki, Stephens]
{Ann Merchant Boesgaard$^1$, Megan C. Novicki$^1$ and Alex Stephens$^1$}

\affiliation{$^1$Institute for Astronomy, University of Hawaii, 2680 Woodlawn
Drive, Honolulu, HI 96822 USA
\break email: boes@ifa.hawaii.edu, mnovicki@ifa.hawaii.edu,
Alex\_C\_Stephens@yahoo.com}

\pubyear{2005}
\volume{228}  
\pagerange{1--7}
\date{?? and in revised form ??}
\jname{From Lithium to Uranium: Elemental Tracers of Early Cosmic Evolution}
\editors{V. Hill, P. Fran\c{c}ois \& F. Primas, eds.}
\begin{document}

\maketitle

\begin{abstract}
Lithium abundances have been determined in more than 100 metal-poor halo stars
both in the field and in clusters.  From these data we find trends of Li with
both temperature and metallicity and a real dispersion in Li abundances in the
Spite Li plateau.  We attribute this dispersion primarily to Li depletion
(presumably due to extra mixing induced by stellar rotation) and to Galactic
chemical evolution.  We derive a primordial Li of 2.44 $\pm$0.18 for A(Li)$_p$
= log N(Li/H) + 12.00.  This agrees with the Li abundances predicted by the
$WMAP$ results.  For stars cooler than the Li plateau we have evidence that Li
depletion sets in at hotter temperatures for the higher metallicity stars than
for the low-metal stars.  This is the opposite sense of predictions from
stellar models.  The smooth transition of the Li content from the Li plateau
stars to the cool stars adds weight to the inference of Li depletion in the
plateau stars.  \keywords{stars: abundances, stars: atmospheres, stars:
kinematics, stars: late-type, subdwarfs, globular clusters, cosmology:
observations}

\end{abstract}

\firstsection 

\section{Introduction}

Since the first publication by \cite{Spite82} about the apparent constancy of
lithium (Li) in low-metallicity stars and the implication that the Li was
produced in the Big Bang, there has been an avalanche of research papers on
this topic.  Here we present high-resolution spectroscopic observations of Li
in over 100 metal-poor halo dwarf and turn-off stars made with the Keck I
telescope and HIRES (\cite{Vogt94}).  There are three inter-twined projects
related to the determination of the value of primordial Li and Li depletion in
old halo stars.  One data set of 38 stars is focussed on the determination of
Li in very low metallicity stars (\cite{nov05}).  Another is the study of Li
in 55 halo dwarfs on extreme orbits which contains a large subset of stars
cooler than 5700 K to examine Li depletion (\cite{BSD05}).  The third is an
investigation of Li in 16 similar turn-off stars in four globular clusters:
M5, M13, M71 and M92 (\cite{BSKD00} and in preparation).

\section{Observational Data}

All of the observations are echelle spectra from HIRES with a spectral
resolution of $\sim$45,000.  The set of the ``very-low-metallicity'' stars
have ratios of signal-to-noise per pixel (S/N) of 100 to 535, with a median of
245.  For the dataset of 55 stars on ``extreme orbits'' the S/N ranges from 70
to 700 with a median of 140.  The 16 stars in the globular clusters are all
near $V$ = 18 so multi-hour exposures were needed to achieve S/N ratios of
only 40 to 60 with a median of 53.

In Figure 1 we show spectrum synthesis fits for Li for some of the stars in
the ``very-low-metallicity'' and the ``extreme-orbits'' data sets.  The
stellar parameters were determined from multiple photometric indices for the
``very-low-metallicity'' set (see \cite{nov05}) and spectroscopically for the
``extreme-orbits'' set (see \cite{SB02}).  Model atmospheres were
interpolated in the Kurucz grid (\cite{K93}) for use in the synthesis which
was done with MOOG (\cite{S73}; http://verdi.as.utexas.edu/moog.html).

\begin{figure}
\centering
\resizebox{6.5cm}{!}{\includegraphics{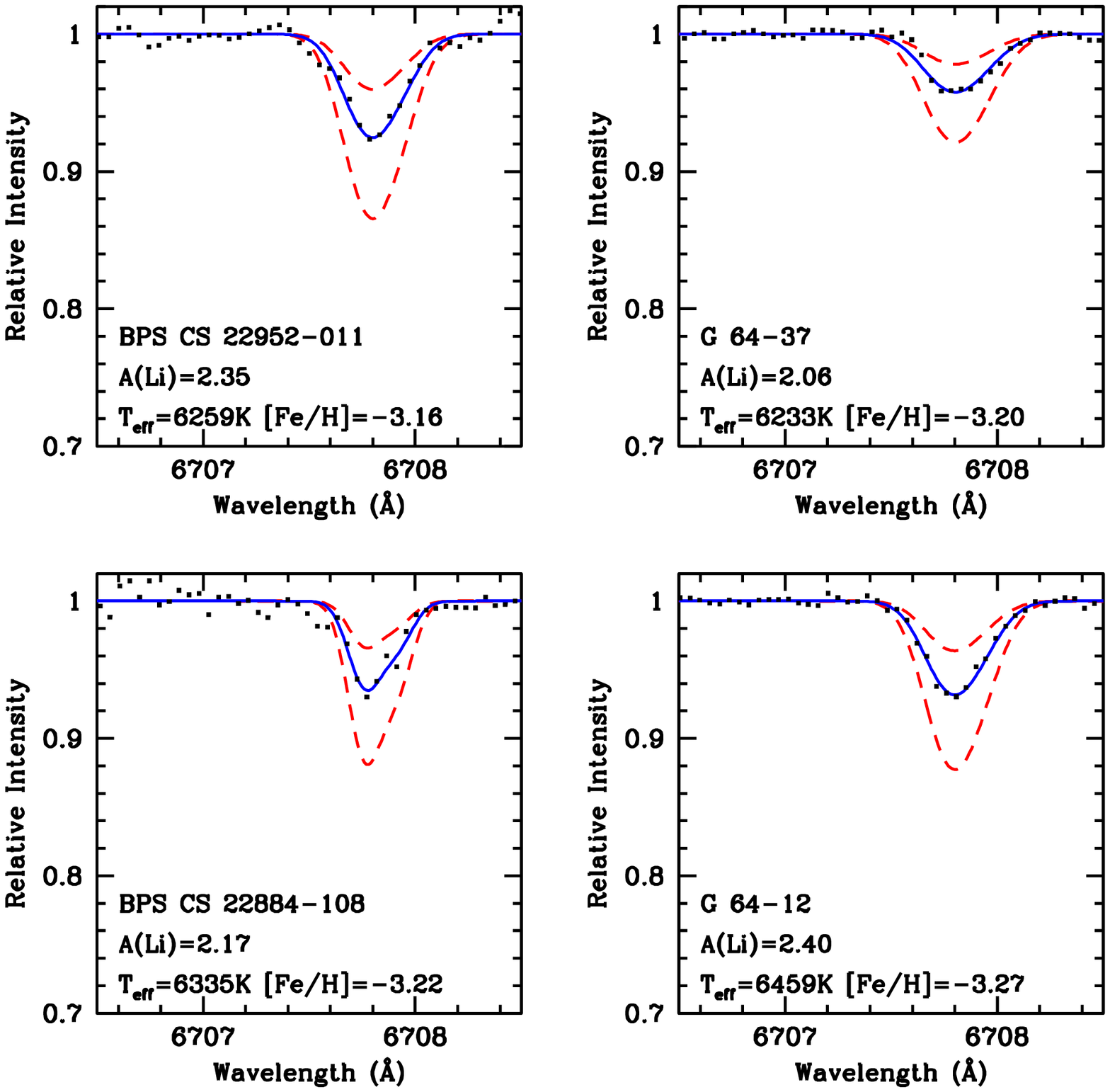} }
\resizebox{6.5cm}{!}{\includegraphics{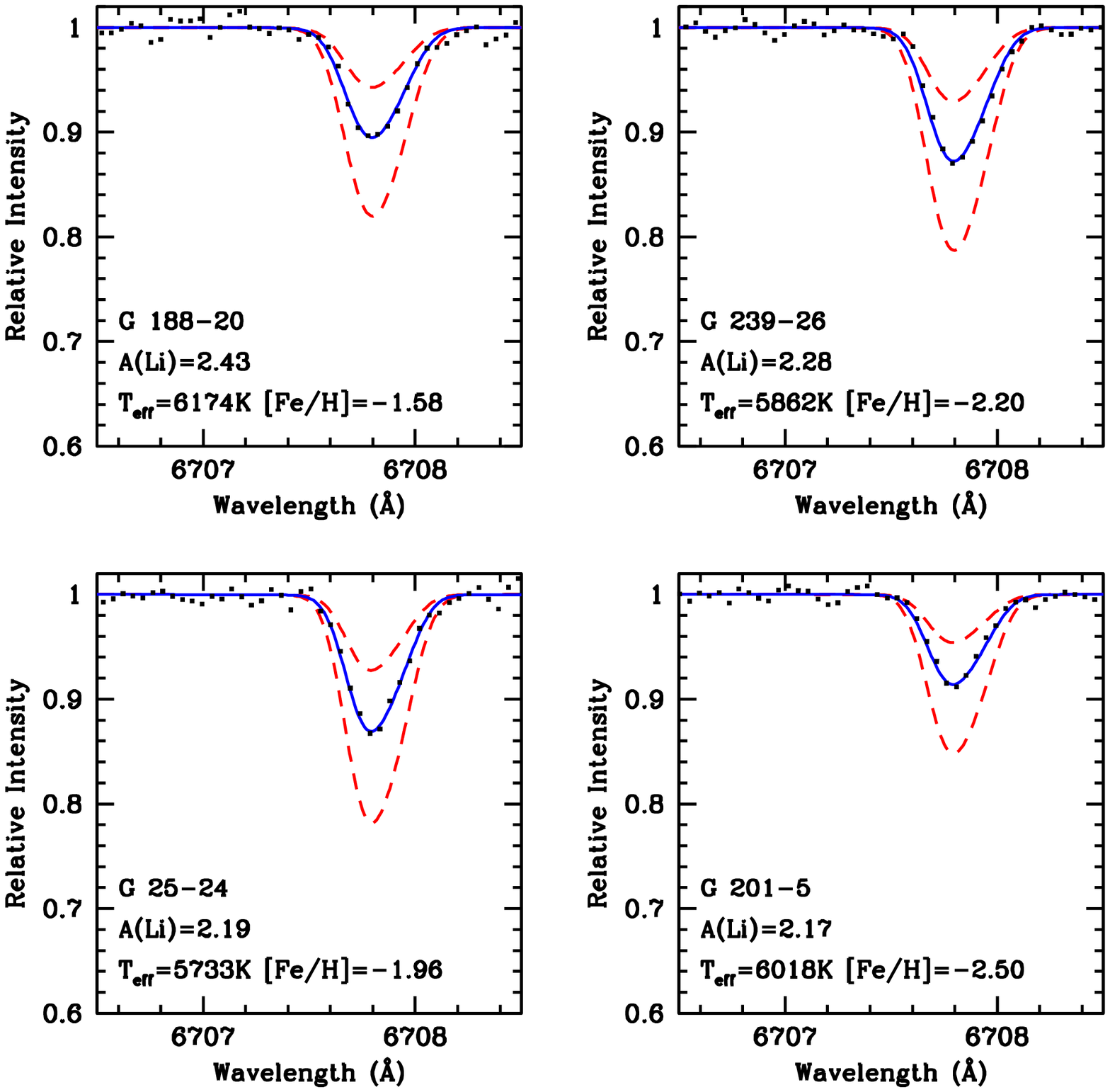} }
\caption[]{Samples of the Li spectrum synthesis fits.  The four stars on the
left are from the very low Fe sample, all with [Fe/H] $<$ $-$3.2, while the
four stars on the right are from the sample of stars on extreme orbits, from
the outer halo or the high halo.  The observations are the black dots.  The
best fit is the solid line and the dashed lines are a factor of two more and a
factor of two less Li.}
\end{figure}

\section{Primordial Lithium}\label{sec:primord}

\subsection{Trends of Li abundance with effective temperature}

The distribution with respect to temperature of the abundance analysis for
A(Li) = log N(Li/H) + 12.00 is shown for the two data sets in Figure 2.  The
higher metallicity and lower log g stars are indicated by different symbols
from the bulk of the stars with low [Fe/H] and high gravities (i.e. halo
dwarfs).  In both samples there is a slight trend for more Li in stars with
T$_{\rm eff}$ $>$ 5700 K.  The Li abundance declines more steeply with
temperature for the cooler stars (T$_{\rm eff}$ $<$ 5700 K).  The right panel
of Figure 2 shows a smooth curve from the hottest to the coolest stars, rather
than an abrupt change at the cool end of the plateau.  As will be discussed
more in $\S$5, the decline in Li, which sets in at about 5700 K, occurs at
hotter temperatures for the more metal-rich stars; the open symbols (higher
metallicity) in the right panel of Figure 2 are shifted toward higher
temperatures relative to the filled symbols.

\begin{figure}
\centering
\resizebox{5.5cm}{!}{\includegraphics{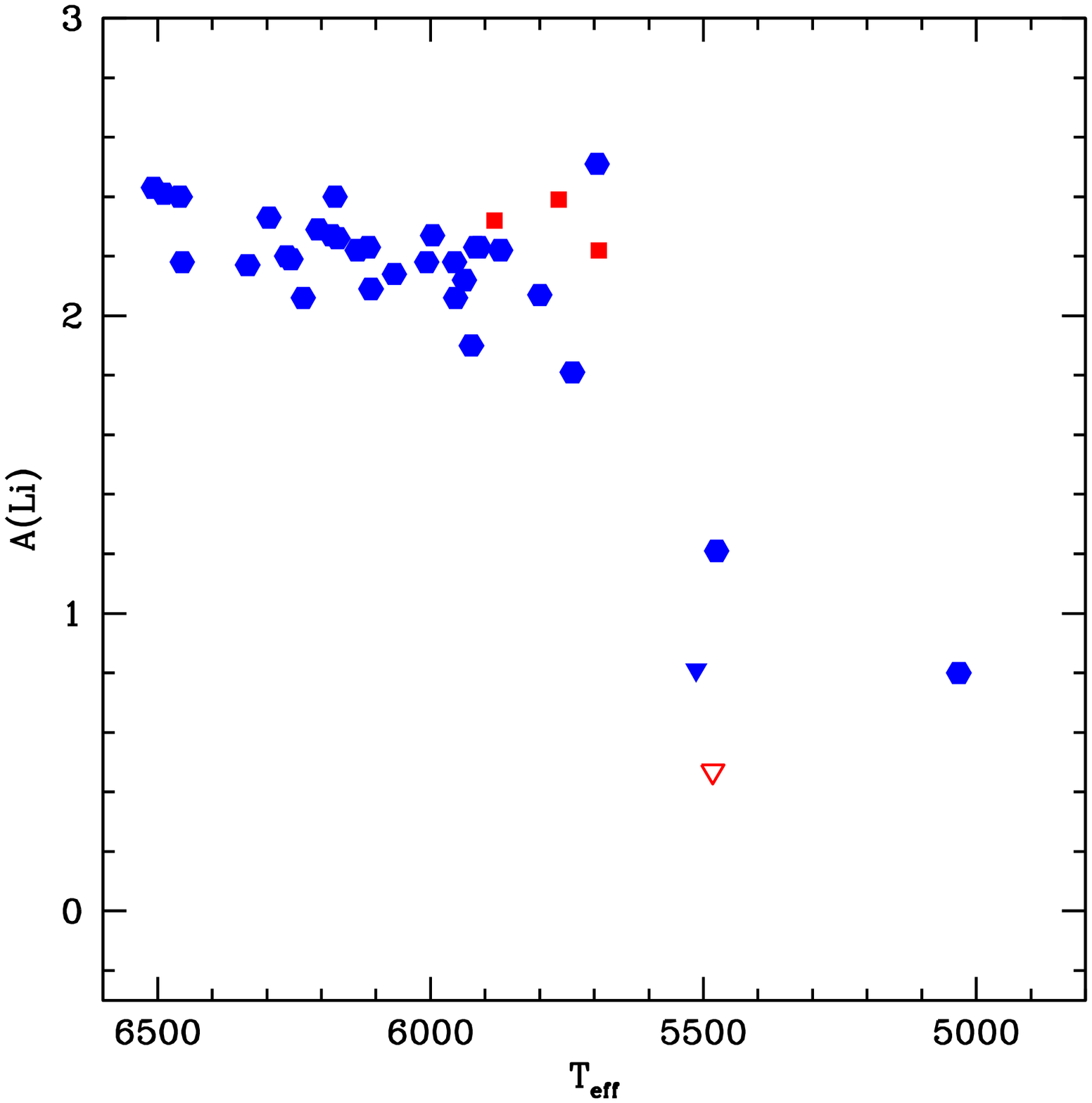} }
\resizebox{5.5cm}{!}{\includegraphics{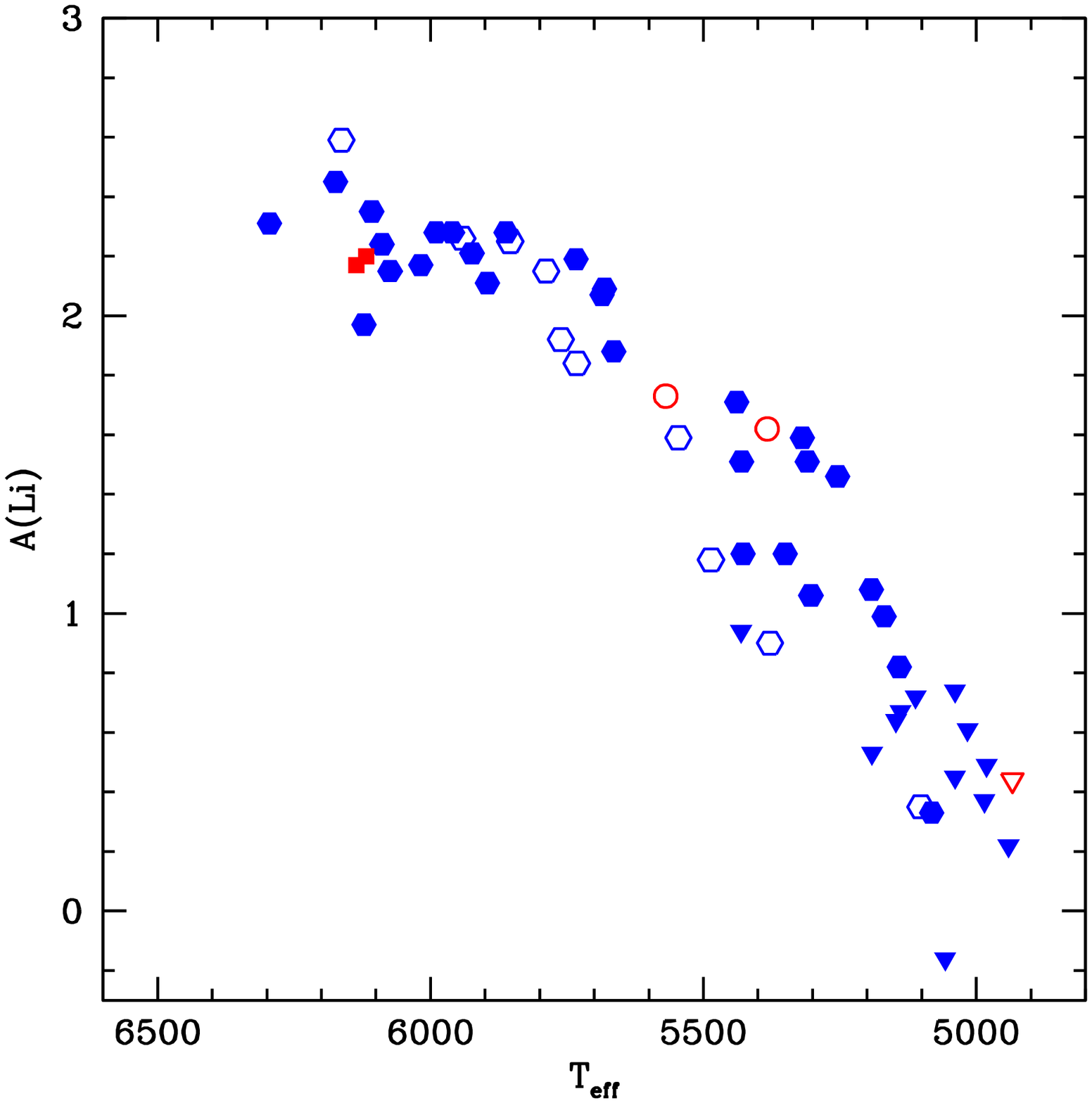} }
\caption[]{The distribution of Li abundances with effective temperatures in
the two star samples: ``very-low-metallicity'' on the left, ``extreme-orbits''
on the right.  The solid symbols are the low metallicity stars ([Fe/H] $<$
$-$1.5).  Open symbols are stars with larger [Fe/H].  Triangles represent upper
limits on A(Li).  Small squares are stars with lower log g ($\sim$3.4-3.5.)}
\end{figure}

\subsection{Trends of Li abundance with Fe and other elements}

The sample of ``extreme-orbits'' stars contains 14 which are on the Li
plateau.  A rather complete chemical profile has been determined for those 55
stars in \cite{SB02}, including Li, Na, Mg, Ca, Ti, Si, Cr, Fe, Ni, Y and Ba.
For the 14 low-metallicity ([Fe/H] $<$ $-$1.5) stars with T$_{\rm eff}$ $>$
5700 K we find trends of increasing Fe-peak abundances with increasing A(Li).
Examples of [Fe/H] and [Cr/H] are found in Figure 3.  The slopes of the
relationships are both 0.18 $\pm$0.04.  For the $\alpha$-elements, [Mg/H],
[Ca/H], and [Ti/H], the slopes with A(Li) are slightly steeper at 0.22
$\pm$0.05, 0.20 $\pm$0.04, and 0.20 $\pm$0.04, respectively.  The slopes are
steeper presumably due to the increasing [$\alpha$/Fe] with decreasing [Fe/H].
The neutron-capture element, [Ba/H], also shows a trend with A(Li), but with a
shallower slope of 0.13 $\pm$0.03.  For more details see \cite{BSD05}.

The larger sample of Li plateau stars that also includes our
very-low-metallicity sample and data from the literature (scaled to match
ours) shows a trend with [Fe/H].  See the left panel of Figure 4.

\begin{figure}[h]
\centering
\resizebox{5.5cm}{!}{\includegraphics{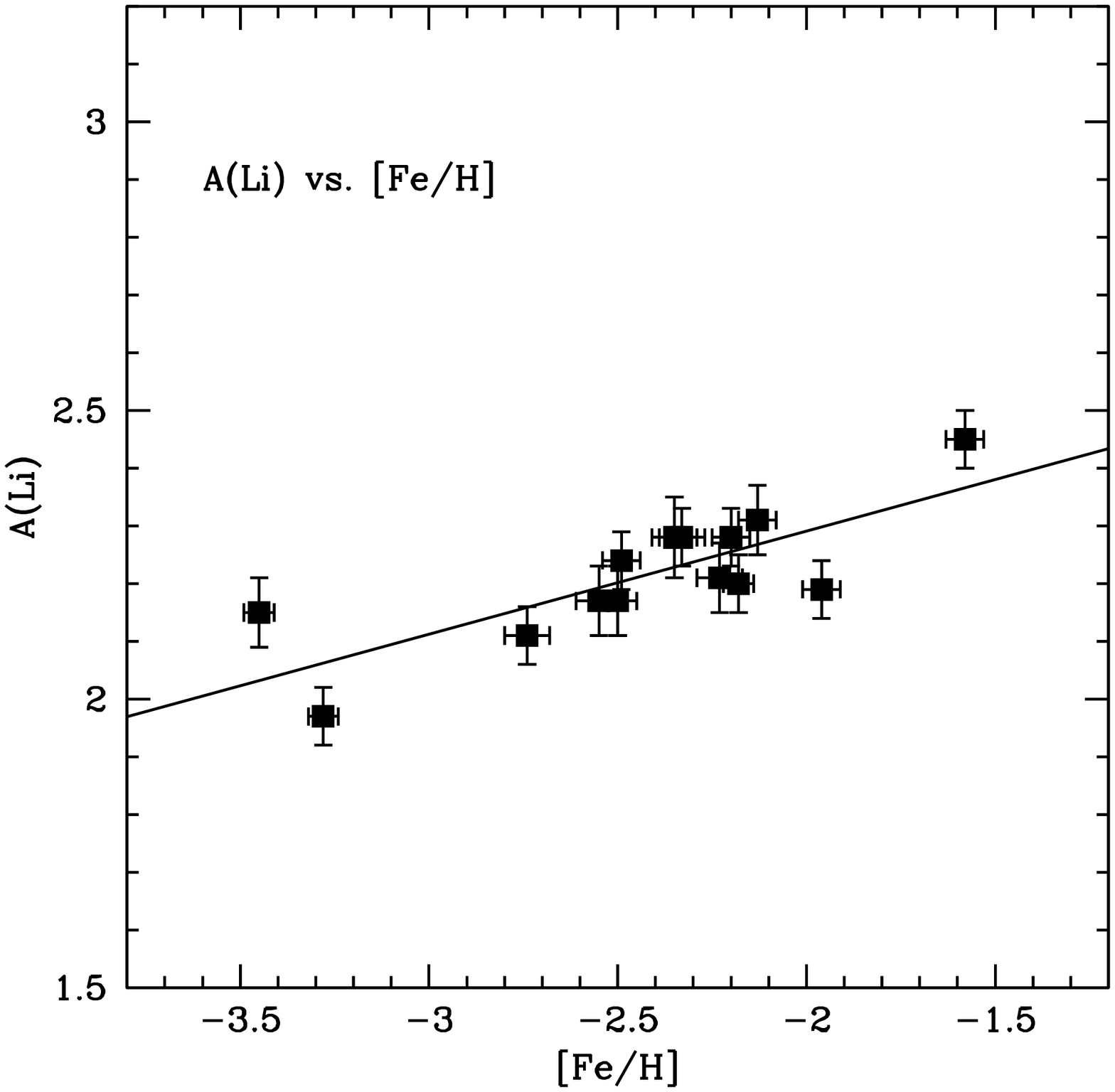} }
\resizebox{5.5cm}{!}{\includegraphics{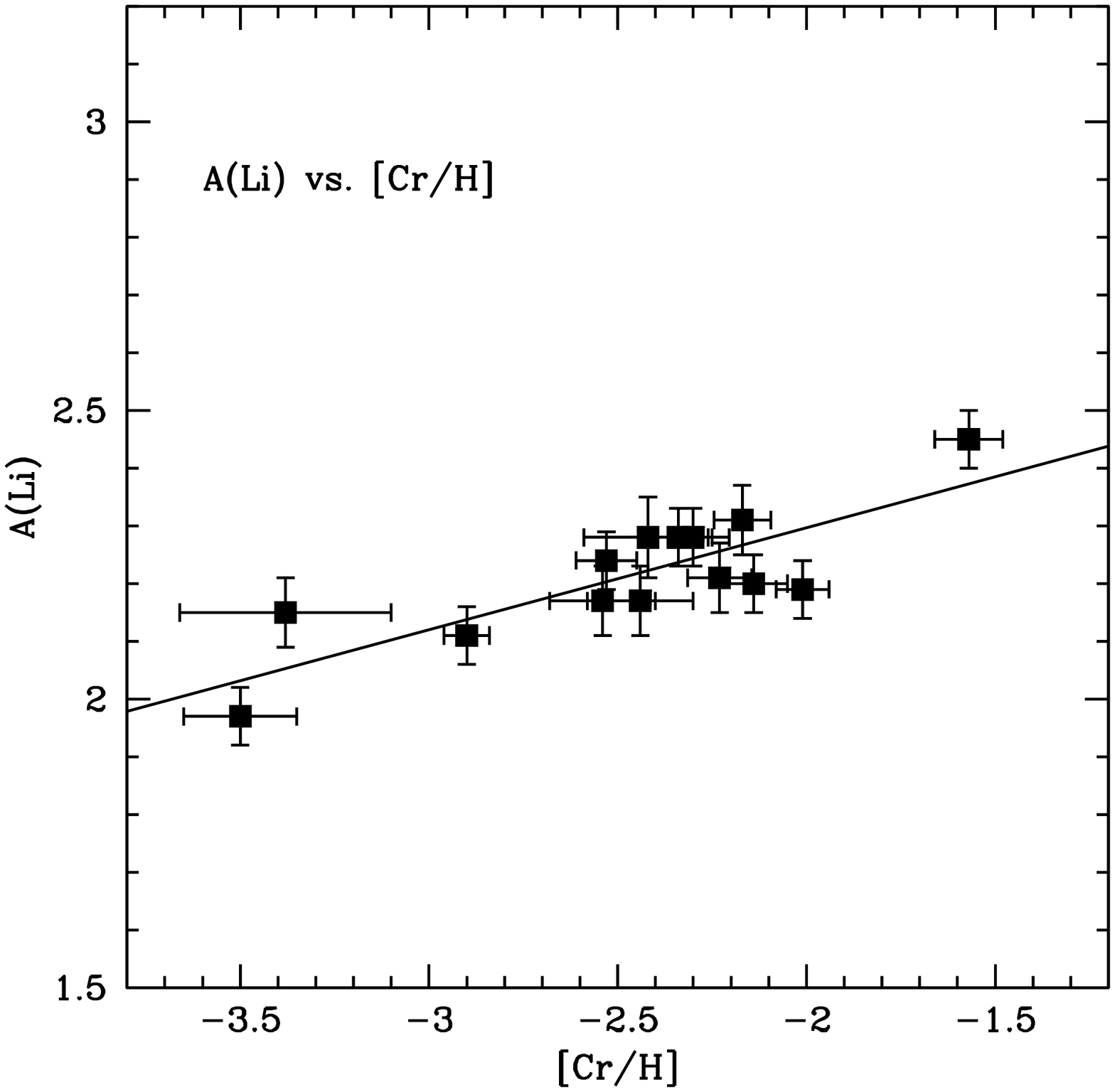} }
\caption[]{The Li abundances in the 14 Li plateau stars from the
``extreme-orbits'' sample are plotted against the Fe-peak element abundances,
[Fe/H] and [Cr/H].  The trends are similar and the slopes are 0.18 $\pm$0.04
for both.}
\end{figure}

\subsection{From observed Li to primordial Li}

Our observations, coupled with those in the literature, show dependencies of
A(Li) on both temperature (Figure 2) and [Fe/H] (Figure 3 and 4-left).  We
have done a bivariate fit to account for the trends with temperature and
metallicity simultaneously for 116 stars: A(Li) = 0.1451 ($\pm$0.0009) [Fe/H]
+ 0.0315 ($\pm$0.0002) T$_{\rm eff}$/100 K.

The errors in the bivariate fit are not truly gaussian.  The difference
between the observed data and the prediction from the bivariate fit show some
points at the high end of the distribution that are not well matched by the
best-fitting gaussian.  We conclude that there is intrinsic dispersion in the
data and that there has been some Li depletion from mixing below the
convection zone.  The level of such depletion has been evaluated at 0.2-0.4
dex by \cite{Pin02}.  We have made corrections for this depletion (+0.3 dex)
and Galactic chemical evolution ($-$0.11 dex) and other small contributors,
following \cite{Ryan02}.  We deduce:
{\bf A(Li)$_p$ = 2.44 $\pm$ 0.18.}
Coc et al.(2004) have found that the predictions from the {\it Wilkinson
Microwave Anisotropy Probe} ($WMAP$) with standard Big Bang nucleosynthesis
result in a value of A(Li)$_p$ = 2.62 $\pm$0.05.  Within the errors our
result agrees with the predictions for A(Li)$_p$ from $WMAP$.

\begin{figure}
\centering
\resizebox{6.5cm}{!}{\includegraphics{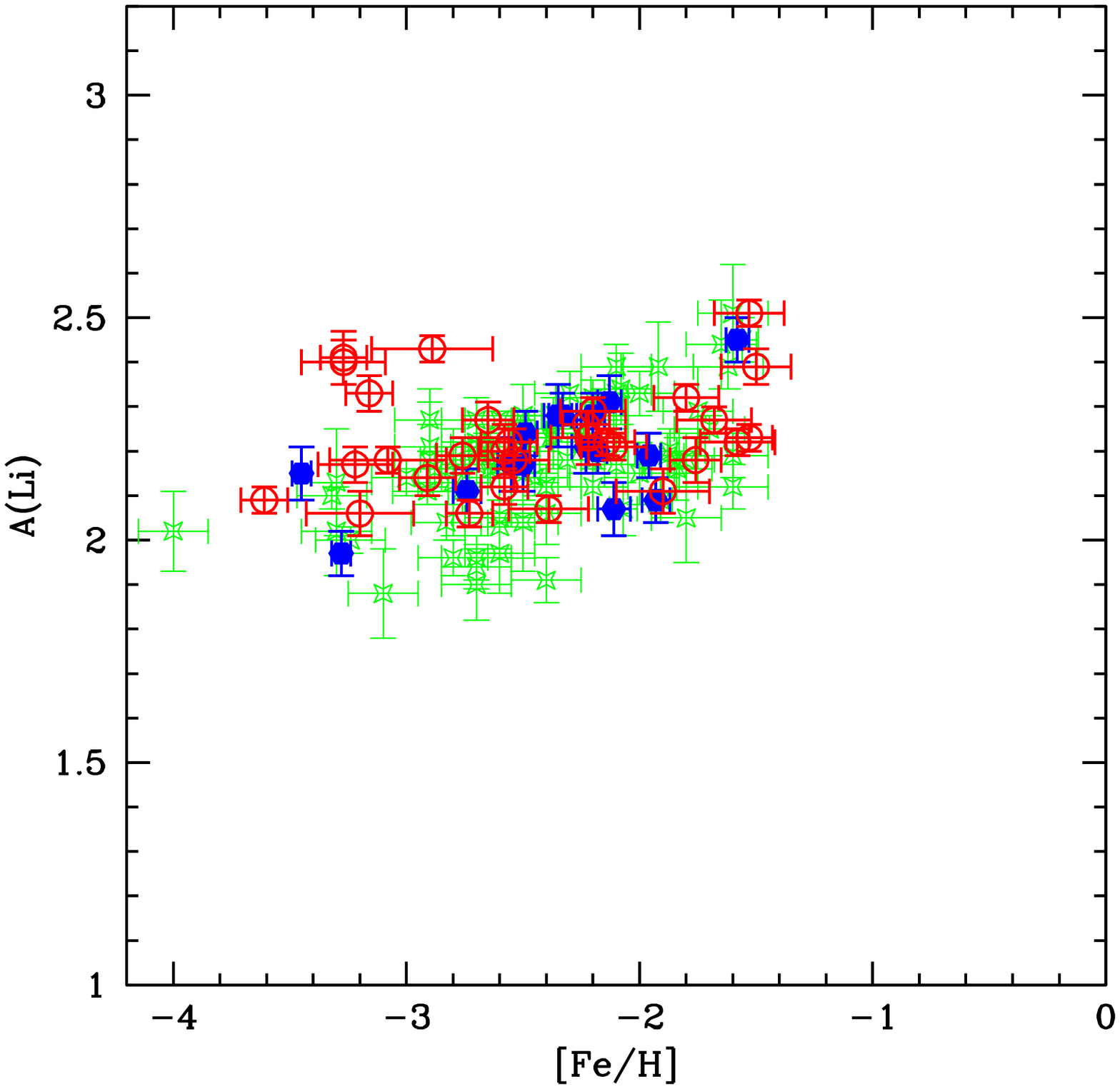} }
\resizebox{6.5cm}{!}{\includegraphics{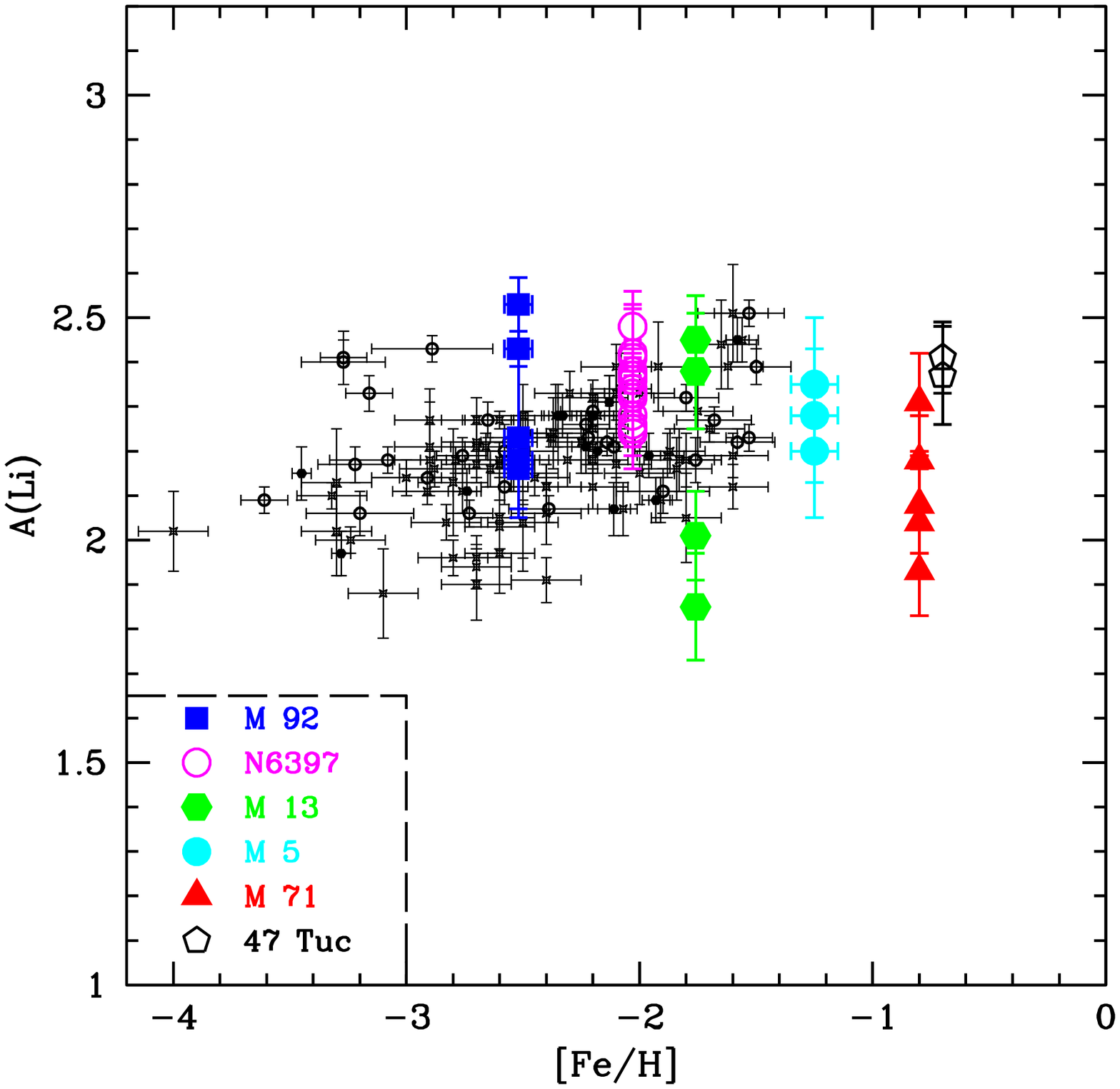} }
\caption[]{Left panel: The Li abundances in the plateau stars as a function of
[Fe/H] for our ``very-low-metallicity'' sample (open squares) and our
``extreme-orbits'' sample (filled squares) with values from the literature on
the same scale (small crosses).  Right panel: The Li abundances for the halo
field stars (small circles) from the left panel, and those from the globular
cluster turn-off stars (symbols indicated in the lower left corner).}
\end{figure}

\section{Li in Turn-off Stars in Globular Clusters}

We have selected stars near the turn-off in four globular clusters which have
as close to the same temperature as possible to study the Li content in these
old stellar populations.  Within a given cluster the stars observed have the
same temperature, the same luminosity, the same mass, the same age, and the
same composition.  The observations are summarized in Table 1.  All the
turn-off stars have temperatures of 5800 - 5900 K.

\begin{table}[h]
\def~{\hphantom{0}}
  \begin{center}
  \caption{Observations of Li in Turn-off Stars in Globular Clusters}
  \label{tab:glob}
  \begin{tabular}{lcccc}\hline
     Cluster  & $[Fe/H]$ & \# Stars & $V$  &  S/N \\\hline
       M 92   & $-$2.52  &    4 & 18.0 & 40 \\
       M 13   & $-$1.76  &    4 & 17.9 & 60 \\
       M 5~   & $-$1.25  &    3 & 18.0 & 57 \\
       M 71   & $-$0.80  &    5 & 17.7 & 55 \\\hline
  \end{tabular}
 \end{center}
\end{table}

There is a range in A(Li) in these 16 stars from 1.85 to 2.53 dex, or almost a
factor of 5.  Observations of turn-off stars in NGC 6397 by \cite{Bon02} and
in 47 Tuc by \cite{PM97} do not show such a large spread.  The range in M13
alone is a factor of 4.  We show the Li spectrum synthesis fits for two of the
M13 stars in Figure 5.  The Li line strengths and Li abundances in these two
otherwise identical stars are clearly different.  Even in M71 at [Fe/H] =
$-$0.80, the spread is a factor of $\sim$2.5.

\begin{figure}[h]
\centering
\resizebox{4.5cm}{!}{\includegraphics{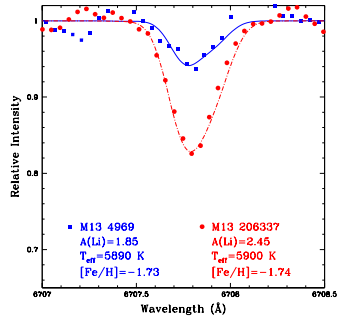} }
\caption[]{The Li line and best fit synthesis for two turn-off stars in the
globular cluster M13.  The solid line (synthesis) and filled squares
(observations) are for star \#4969 and the dot-dashed line and filled circles
are for star \#206337.  The two stars are identical in temperature,
luminosity, age, mass and composition, except for the Li abundances which
differ by a factor of 4.}
\end{figure}

In Figure 4 (right) we show the Li results for the halo field stars and the
globular cluster stars.  It is possible that the spread revealed in the
cluster stars results from depletion due to rotationally-induced mixing;
the spread could be larger in the cluster stars because the range in initial
angular momenta could be greater in the cluster-forming environment than in
the field.  All the stars have spun-down and are now slow rotators, but those
that were initially rotating more rapidly would be the ones that are more
Li-depleted now.

\section{Li Depletion in Cool Halo Dwarfs}

The right panel in Figure 2 shows some 22 stars cooler than 5800 K with log g
$>$ 3.7 and with detectable Li.  When we subdivide these stars into three
groups with similar Fe abundances, we see three distinct, but similar,
patterns of depletion.  The left panel of Figure 6 show the trends of A(Li)
with temperature for the three groups, and shows the least squares fits
through those points (one fit for the higher metallicity stars, 2 for the two
lower ones.)  (One star at [5254,1.46] was excluded from the fit for the
middle group; if its [Fe/H], at $-$1.47 $\pm$0.05, were smaller by 0.11 dex,
it would be part of the lowest metallicity group.)  It can be seen that the
decline in Li sets in at higher temperatures for the higher metallicity stars,
the mid-range metallicity stars start at lower temperatures, followed by the
lowest metallicity at the coolest temperatures.  The right panel shows the the
theoretical predictions from the Yale ``standard'' models (\cite{RD98}).
Those predictions are in the opposite sense of the observations; they predict
more Li depletion, not less, in the most metal-poor stars, and they predict a
steeper decline than is observed.

\begin{figure}
\centering
\resizebox{5.5cm}{!}{\includegraphics{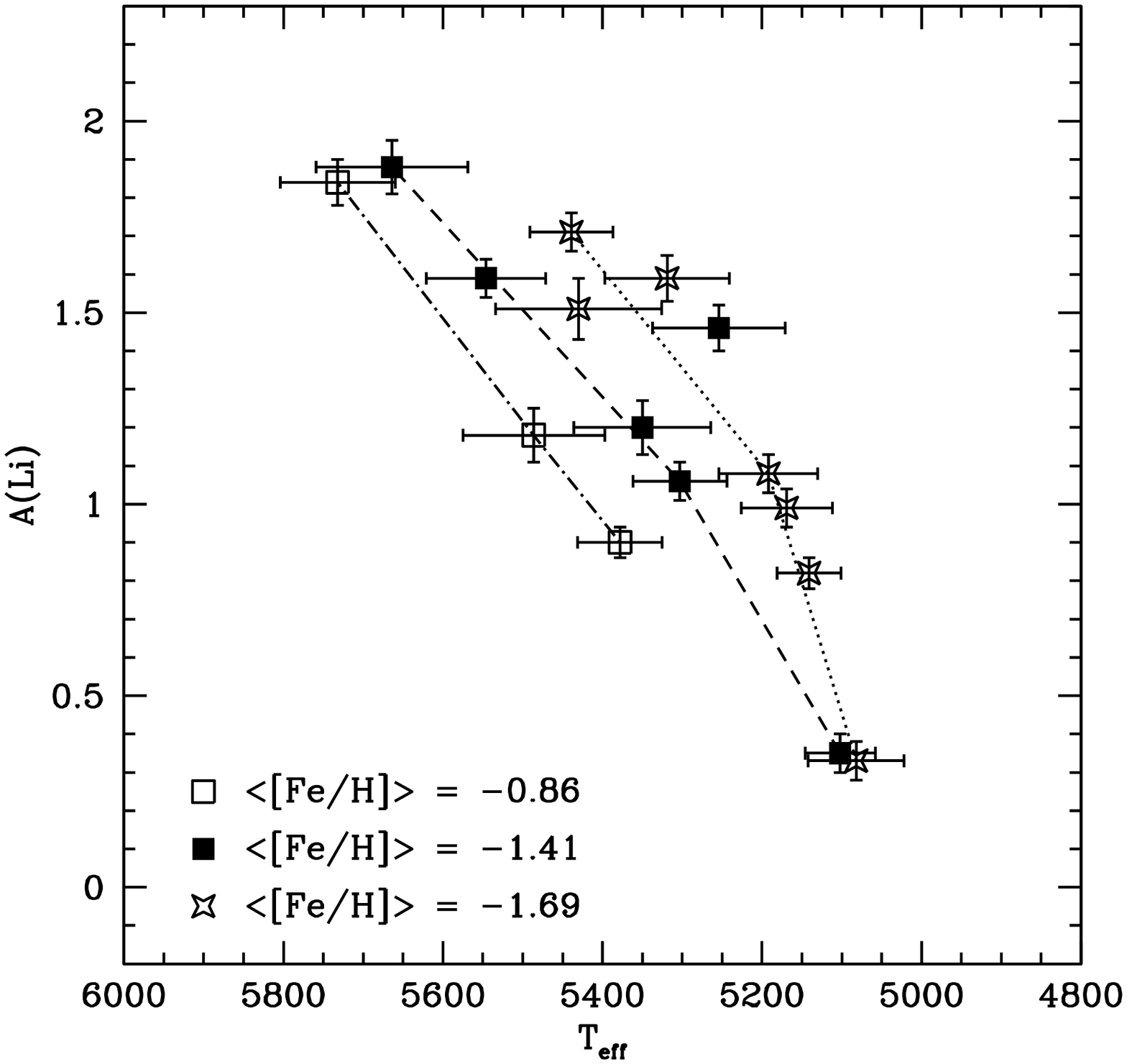} }
\resizebox{5.5cm}{!}{\includegraphics{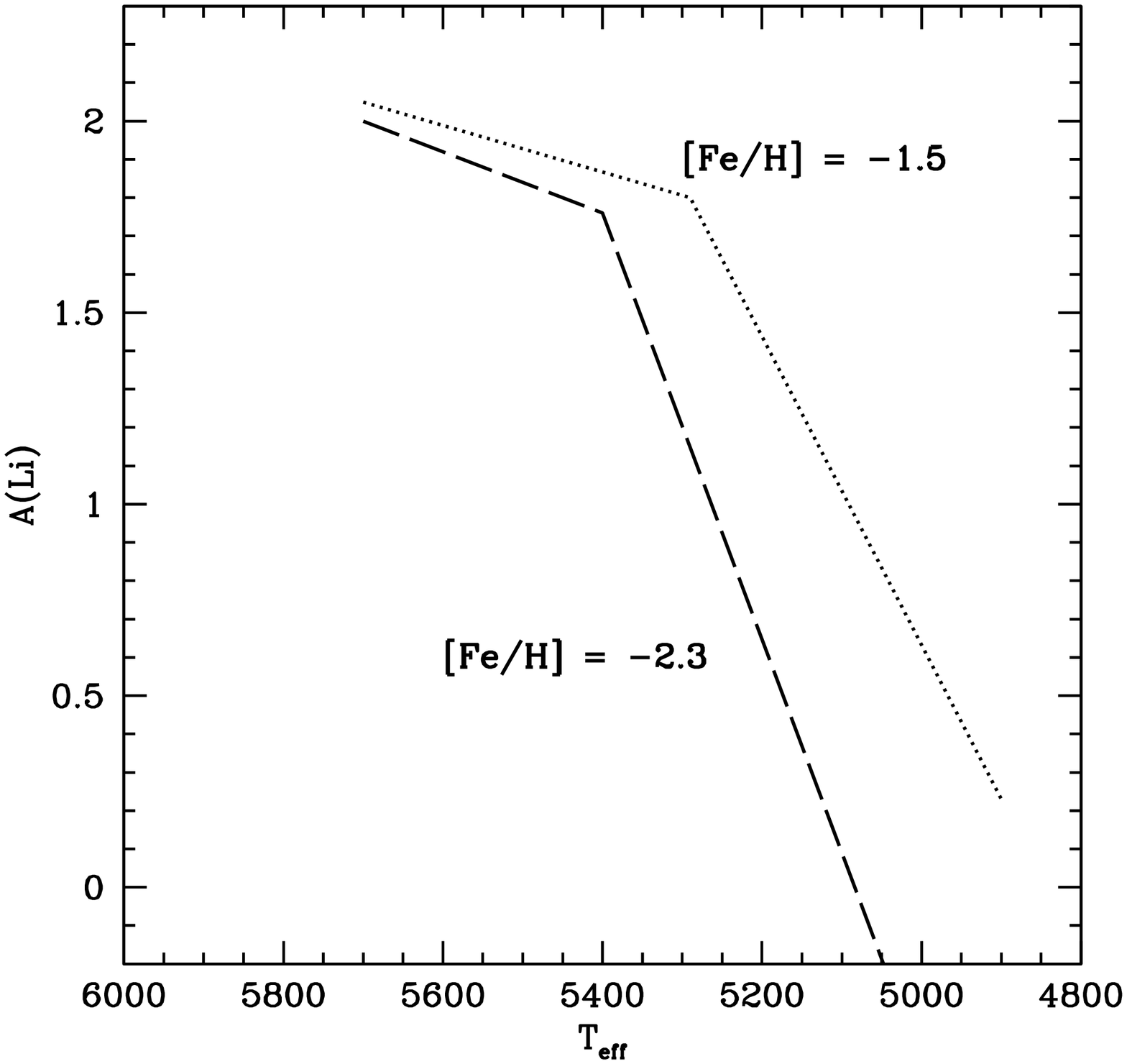} }
\caption[]{Left panel: Li depletion in cool halo dwarfs.  Three metallicity
groupings are shown with [Fe/H] = $-$0.70 to $-$1.03; $-$1.24 to $-$1.47; and
$-$1.58 to $-$1.90.  The Li depletion begins at higher temperatures for the
higher metallicity stars.  Right panel: Standard ``Yale'' model predictions
for Li depletion in 16.5 Gyr stars plotted on the same scale.  The predictions
are in the opposite sense of the observations.}
\end{figure}

\section{Conclusions}\label{sec:concl}

We find evidence for a dispersion in the Spite Li plateau.  This appears to
result both from Li depletion due to rotationally-induced mixing and from a
small increase in Li due to Galactic chemical evolution.  Our value for
primordial Li, A(Li)$_p$, is 2.44 $\pm$0.18 after taking into account various
effects on the observed Li; this agrees with the predictions from the results
of ${\it WMAP}$ of 2.62 $\pm$0.05.  For the globular cluster turn-off stars
our Li abundances show a spread at a given temperature and metallicity of up
to a factor of 4.  This spread could be due to different degrees of Li
depletion in these stars, which possibly results from rotationally-induced
mixing and a range in initial angular momentum in otherwise similar cluster
stars.  The original rapid rotators would have lost more Li than the slow
rotators although all are rotating slowly now.

We have measured Li depletion in cool metal-poor stars and find it to be a
function of metallicity at a given temperature.  There is a range of a factor
of 4 at 5400 K, for example.  The steep decline in Li begins at higher
temperatures for stars with higher values of [Fe/H] than for those with lower
[Fe/H].  The theoretical predictions are in the opposite sense of the
observations and predict more severe depletion.

\begin{acknowledgments}
This work has been supported by NSF grants to AMB: AST0097945 and AST0097955.
\end{acknowledgments}

\end{document}